\documentclass[useAMS, usenatbib, preprint, 12pt]{aastex}
\usepackage{cite, natbib}
\usepackage{float}
\usepackage{epsfig}
\usepackage{cases}
\usepackage[section]{placeins}
\usepackage{graphicx, subfigure}
\usepackage{color}

\newcommand{\article}{article}
\newcommand{\nGO}{4923}
\newcommand{\uHz}{$\mu$Hz}
\newcommand{\oxford}{1}
\newcommand{\nyu}{2}
\newcommand{\cfa}{3}

\newcommand{\rpc}{201133037}
\newcommand{\rpcc}{201142043}

\begin{document}

\title{Systematics-insensitive periodic signal search with K2}

\author{%
   Ruth Angus\altaffilmark{\oxford, \cfa}
   Daniel Foreman-Mackey\altaffilmark{\nyu}
   \& John A. Johnson\altaffilmark{\cfa}
}

\altaffiltext{\oxford}{Subdepartment of Astrophysics, University of Oxford, OX1 3RH, UK, ruthangus@gmail.com}
\altaffiltext{\nyu}{Center for Cosmology and Particle Physics, New York University, NY, USA}
\altaffiltext{\cfa}{Harvard-Smithsonian Center for Astrophysics, 60 Garden St.,
Cambridge, MA, USA}

\begin{abstract}

From pulsating stars to transiting exoplanets, the search for periodic signals
in {\it K2} data, {\it Kepler's} 2-wheeled extension, is relevant to a long
list of scientific goals.
Systematics affecting {\it K2} light curves due to the decreased
spacecraft pointing precision inhibit the easy extraction of periodic signals
from the data.
We here develop a method for producing periodograms of K2 light curves that
are insensitive to pointing-induced systematics; the Systematics-Insensitive
Periodogram (SIP).
Traditional sine-fitting periodograms use a generative model to find the
frequency of a sinusoid that best describes the data.
We extend this principle by including systematic trends, based on a set of
`Eigen light curves', following \citet{Foreman-Mackey2015}, in our generative
model as well as a sum of sine and cosine functions over a grid of
frequencies.
Using this method we are able to produce periodograms with vastly reduced
systematic features.
The quality of the resulting periodograms are such that we can recover
acoustic oscillations in giant stars and measure stellar rotation periods
without the need for any detrending.
The algorithm is also applicable to the detection of other periodic phenomena
such as variable stars, eclipsing binaries and short-period exoplanet
candidates.
The SIP code is available at \url{https://github.com/RuthAngus/SIPK2}.

\end{abstract}


\section{Introduction}
\label{Introduction}

The excellent precision achieved by the original {\it Kepler} mission relied
on extremely precise pointing, for which three reaction wheels were required.
After the failure of one of these wheels, the {\it Kepler} team devised a new
pointing scheme in which the spacecraft is stabilized by the Solar wind for
ecliptic plane viewing zones \citep{Howell2014}.
In this configuration the spacecraft is able to maintain an unstable
equilibrium, with the two functioning reaction wheels controlling pitch and
yaw whilst the spacecraft slowly rolls about the boresight.
The spacecraft fires its thrusters once every $\sim$ 6 hours
\citep[][hereafter VJ14]{Vanderburg2014} to correct for
this slow drift and, as stars move across pixels with different sensitivities,
their flux varies.
The extraction of high-precision photometry from {\it K2} target pixel files,
despite the reduced pointing precision, is a requirement for many fields of
research and several methods for the extraction and detrending of {\it K2}
light curves have already been developed.
For example, VJ14 and \citet{Crossfield2015}
use simple aperture photometry and correct the light curve of each star
individually and \citet{Aigrain2015} use a Gaussian process to model the
non-linear dependence of stellar flux on the roll angle of the telescope.

\begin{figure}[p]
\begin{center}
\includegraphics[width=6in, clip=true]{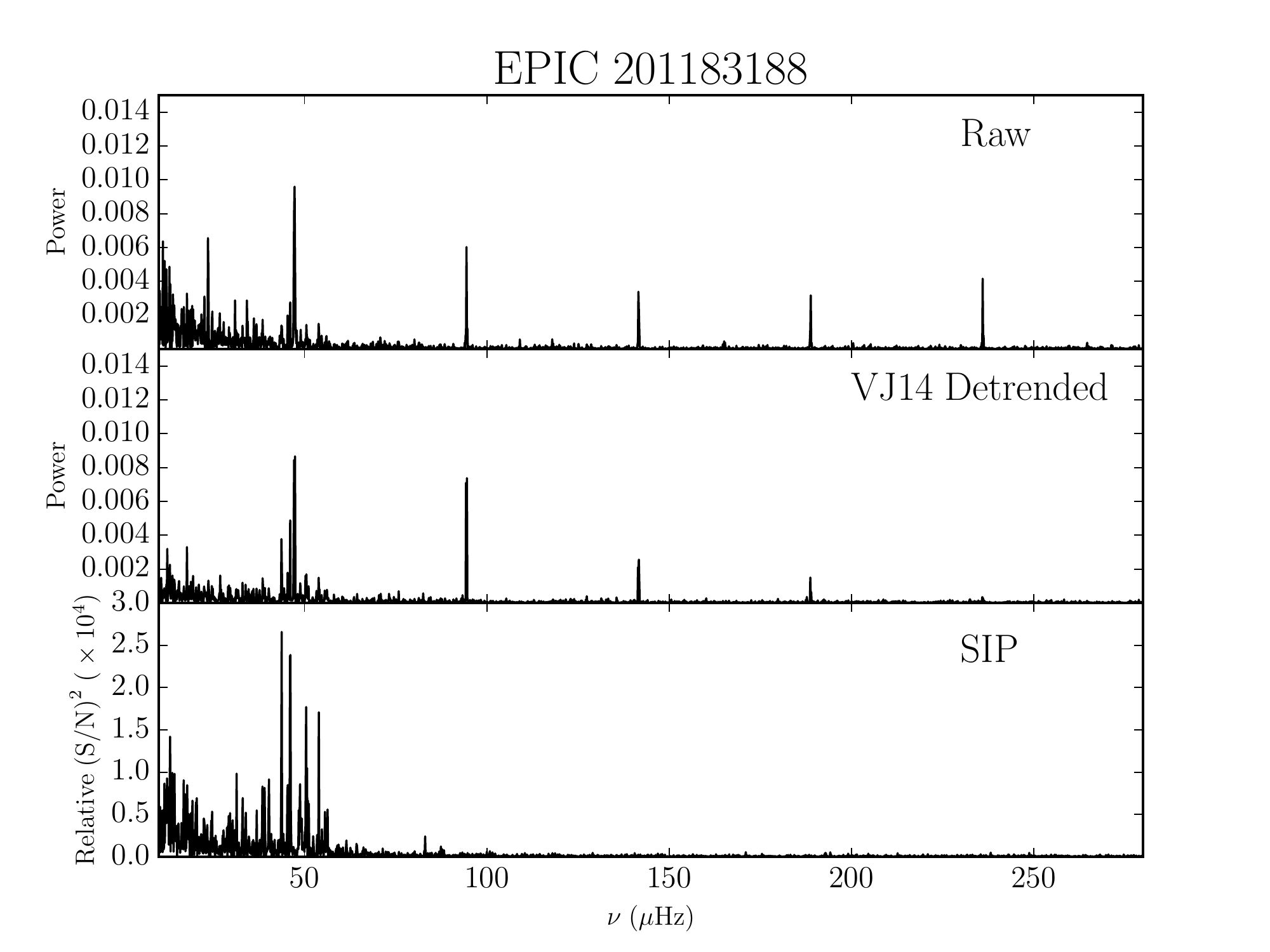}
\caption{LS periodograms of the raw (Top) and VJ14--detrended (middle) {\it K2}
	 light curves of EPIC 201183188.
	 The bottom panel shows the SIP for this target.
	 Peaks at $\sim$ 47 $\mu$Hz and its harmonics produced by the regular
	 spacecraft thruster fires are still present in the LS periodogram of
	 the detrended data, but do not appear in the SIP.}
\label{fig:raw}
\end{center}
\end{figure}

Whilst these methods successfully remove most systematic trends and
produce light curves suitable for exoplanet search and some stellar
variability studies, residual systematics can still affect the light curves on
timescales relevent to asteroseismology and stellar rotation.
In particular, the $\sim$ 6 hour thruster firing signal may still appear
with high power in the periodograms of these detrended light curves
(see figure \ref{fig:raw}).
A detrending method for {\it K2} light curves, specifically intended for the
asteroseismic analysis of giant stars has been developed by \citet{Lund2015},
in which the systematics due to roll are corrected, again on
a star-by-star basis and any remaining periodic signals at 47 $\mu$Hz (6 hour
period) or its harmonics are removed by prewhitening.
The method developed here, the Systematics-Insenstive Periodogram (SIP)
produces periodograms of {\it K2} light curves without the need for detrending
or prewhitening.

\subsection{Asteroseismology}

As well as providing data that have lead to the discovery thousands of
exoplanets, the original {\it Kepler} mission revolutionized many fields of
stellar astronomy, particularly asteroseismology.
Fundamental stellar parameters---in some cases, extremely precise ones---can
be calculated for {\it Kepler} asteroseismic stars from the power spectra of
their light curves.
Although Sun-like stars oscillate at high frequencies and require
short-cadence observations, pulsations of giant stars lie below the Nyquist
frequency set by the 28.5 minute sampling rate of long cadence {\it Kepler}
data: 283 $\mu$Hz.
Asteroseismic analysis of data from the original {\it Kepler} mission is
traditionally conducted upon detrended light curves.
For short cadence {\it Kepler} data, this detrending method is described in
\citet{Garcia2011}.
Due to the precise pointing of the original {\it Kepler} mission, systematics
present in these light curves, caused by temperature fluctuations and minor
pointing shifts, are relatively low amplitude.

However, this is not the case for {\it K2} light curves: the precision over a
6 hour timescale is estimated to be 4 times worse in {\it K2} data
\citep{Howell2014}, therefore new approaches to the treatment of systematics
are necessary.
Figure \ref{fig:raw} demonstrates the need for careful systematics treatment
of {\it K2} photometry for asteroseismology.
The top panel shows a Lomb-Scargle\footnote{All LS periodograms
produced in this project were produced using the gatspy Python module:
\url{https://github.com/astroML/gatspy/tree/master/gatspy/periodic}} (LS)
periodogram of the raw, simple aperture photometry\footnote{The method used to
extract this photometry is described in \textsection \ref{sec:Method}} of EPIC
201183188, a pulsating giant star.
The large peaks at $\sim$ 47 $\mu$Hz and its harmonics are caused by the
regular thruster fires of the spacecraft.
The bottom panel shows the LS periodogram of this light curve, after
it has been detrended using the method of VJ14.
The large peaks are still present in the detrended light curve.
While this remaining noise source does not interfere with the detection of
high-signal-to-noise transit events for periods greater than $\sim$1 day
\citep{Vanderburg2015}, it does hamper the detection of
smaller signals, particularly on time scales comparable to that of thruster
fires.
These peaks lie in an important region of parameter space for giant star
asteroseismology and could affect the stellar parameters measured for thousands
of giants if not dealt with appropriately.

\subsection{Stellar rotation}

Stellar rotation studies have hugely benefitted from the era of high-precision
space photometry.
Active regions on the surface of rotating stars produce periodic variations
in flux and stellar rotation periods can therefore be measured from
{\it Kepler} light curves.
Stellar rotation is a field of active interest as the rotation period of star
can be used to infer its age via gyrochronology
\citep{Skumanich1972, Barnes2007, Epstein2014, Angus2015}, is thought to be
tied to the stellar magnetic dynamo, and could even reveal dynamical
interations with companion stars or planets
\citep[e.g.][]{Beky2014, Poppenhaeger2014}.
Current methods for measuring rotation periods from {\it Kepler} light curves
include periodogram \citep[e.g.][]{Reinhold2013}, AutoCorrelation Function
(ACF) \citep{McQuillan2013} and wavelet \citep[e.g.][]{Garcia2014} analysis,
or some combination thereof.
Stellar variability is not typically sinusoidal, therefore sine-fitting
periodograms are not perfectly suited to measuring rotation periods
\citep{McQuillan2013}.
For this reason, the ACF method is often favoured over the periodogram method.
However, because autocorrelation is performed directly on detrended light
curves, and cannot be written down as a generative model, it is not possible
use autocorrelation techniques on untreated {\it K2} data.
A quasi-periodic Gaussian process is a much better effective model for stellar
variability than a sinusoid, however we choose to focus on the more generally
applicable (and computationally tractable) sine-wave periodogram, leaving the
Gaussian process periodogram for future consideration.

In this \article\ we focus on the examples of asteroseismology and stellar
rotation, however many other fields of astronomy utilize periodic information
in {\it K2} light curves.
These include studies of eclipsing binaries, variable stars, exoplanets, white
dwarfs and even AGN.
The development of tools for extracting periodic information from {\it K2}
data is essential if it is to be as revolutionary in time-domain
astronomy as the original {\it Kepler} mission was.

In \textsection\ref{sec:Method} we outline the method behind the SIP.
In \textsection\ref{section:rotation} we apply the SIP to real {\it K2} light
curves, using some giant asteroseismic pulsators and rotating stars as test
cases and then provide the results of some simple tests which show exactly
{\it how} `insensitive' the SIP is to systematic features.
Finally, we demonstrate the SIP's usefulness regarding other periodically
varying objects in this section, before presenting our conclusions in
\textsection\ref{sec:conclusions}.

\section{Method}
\label{sec:Method}

The method implemented in this \article\ is an extention of the planet-search
algorithm developed by \citet{Foreman-Mackey2015} (hereafter FM15).
All targets observed by {\it Kepler} move on the CCD in the same way,
therefore the systematics affecting each individual star's light curve have
shared properties.
The FM15 method uses this fact by decomposing the light curves into a set
of `Eigen Light Curves' (ELCs) using Principle Component Analysis (PCA), which
can be used to model any individual star's light curve with very little loss
of information.
This process is similar to the method used to produce PDC-MAP data for the
original {\it Kepler} mission \citep[][]{Stumpe2012, Smith2012}.
The resulting ELCs from campaign 1 can be used to model any campaign 1 {\it
K2} light curve, (campaign 0 ELCs for campaign 0, etc) and specifically, can
model the data in combination with an arbitrary physical model.

In order to construct sets of ELCs for campaigns 0 and 1, FM15 downloaded the
target pixel files for all stars in these two fields.
The position of each star was predicted using the World Coordinate System (WCS)
and 10 circular apertures placed around the star with radii varying from 1 to
5 pixels in steps of 0.5 pixels.
Following the procedure of VJ14, the aperture producing the
light curve with the lowest CDPP within a 6 hour window
\citep{Christiansen2012} was selected\footnote{The simple aperture photometry
light curves for campaigns 0 and 1 are available at
\url{http://bbq.dfm.io/ketu/lightcurves/}}.
PCA was then performed on the full set of targets in order to produce ELCs.

FM15 used 150 of these ELCs, plus a transit model, in order to
search for exoplanet candidates without the need for a `detrending' step.
The likelihood of the data, conditioned on the ELC-plus-transit
model was calculated over a fine grid of periods and transit depths, resulting
in the detection of 36 new exoplanet candidates.
We use a very similar technique to find periodic signals in K2 data.
The primary difference is that we use a sinusoid rather than a transit model.
This model is linear, therefore the likelihood function conditioned on
a specific frequency can be calculated and the systematics model marginalized
over analytically.

Following the notation in FM15, our model for the $k$th star can be written
\begin{equation}
	\mathbf{f_k} = \mathbf{A}\mathbf{w_k} + \mathrm{noise},
\end{equation}
where $\mathbf{f_k}$ is the vector of $N$ flux values,
\begin{equation}
	\mathbf{f_k} = (f_{k,1}, f_{k,2}, f_{k,3}, ..., f_{k,N})^T
\end{equation}
at times
\begin{equation}
	\mathbf{t_k} = (t_1, t_2, t_3, ..., t_N)^T.
\end{equation}
$\mathbf{A}$ is the design matrix:
\begin{eqnarray}
	\mathbf{A} &=& \left (\begin{array}{ccccccc}
	x_{1,1} & x_{2,1} & \cdots & x_{150,1} & 1 & \sin(2\pi\nu t_1) & \cos(2\pi\nu t_1) \\
	x_{1,2} & x_{2,2} & \cdots & x_{150,2} & 1 & \sin(2\pi\nu t_2) & \cos(2\pi\nu t_2\\
    && \vdots &&&\\
	x_{1,N} & x_{2,N} & \cdots & x_{150,N} & 1 & \sin(2\pi\nu t_N) & \cos(2\pi\nu t_N)
\end{array}\right )
\end{eqnarray}
where the $x_{ij}$s are the ELCs\footnote{Campaign 0 and 1 ELCs are
available at \url{http://bbq.dfm.io/ketu/elcs/}}, with $i$ denoting the ELC number
and $j$ the time index.
The design matrix contains the basis functions of the linear model.
The basis functions for the systematic features in the light curves are the ELC
values at each time index, the sine and cosine terms are the basis functions of
the sinusoidal signal of interest, and the `1's describe a linear offset.
Any {\it K2} light curve can be reproduced as a linear combination of these
basis functions.
We are interested in the last two elements of the weight vector: the
coefficients of the sinusoidal signal.
The maximum likelihood solution for the weight vector,
$\mathbf{w}$ is
\begin{equation}
	\mathbf{w_k}^* \gets (\mathbf{A}^T\mathbf{A})^{-1}\mathbf{A}^T\mathbf{f_k}.
\end{equation}

Under this linear model with Gaussian uncertainties, the marginalized
likelihood for the periodic amplitude is a two-dimensional Gaussian with mean
given by the last two elements ($\mathbf{a}$) of $\mathbf{w}*$ and covariance
given by the bottom right two-by-two block ($\mathbf{S_a}$) of
$(\mathbf{A}^T \mathbf{C}^{-1} \mathbf{A})^{-1}$, where the $\mathbf{C}$
matrix contains observational uncertainties on the diagonal.
These uncertainties are estimated as $1.48 \times$ the Median Absolute
Deviation (MAD), following \citet{Aigrain2015}.
Therefore, the signal-to-noise ratio, $S/N$ of the amplitude measurement is
$\sqrt{\mathbf{a}^T \mathbf{S_a}^{-1} \mathbf{a}}$.
The $(S/N)^2$ is calculated over a grid of frequencies to produce an
SIP.
The $(S/N)$ operation takes into account the goodness of fit, i.e. if the
amplitude of the sinusoid at a given frequency is not well constrained,
it is penalized.

\section{Application to real light curves}
\label{section:rotation}

An example LS periodogram of the raw {\it K2} photometry
for giant star, EPIC 201183188 is shown in figure \ref{fig:raw}.
Peaks appearing at 47 $\mu$Hz and its harmonics are produced by the regular
$\sim$ 6 hour thruster fires that repoint the spacecraft.
These peaks are also present in periodograms of the VJ14 detrended light
curves.
The presence of systematic signals at these timescales are problematic for
asteroseismic analysis since they lie in a region of frequency space
that is often populated by giant asteroseismic modes.
It is possible to remove these signals by `prewhitening' the data, i.e.
subtracting a sinusoid of that frequency from the data, however this process
will artificially supress all signals, both systematic and astrophysical, at
that frequency.
The SIP method eliminates the necessity for any such procedure.
The bottom panel of figure \ref{fig:raw} shows the SIP for the same star,
demonstrating the ability of the SIP method to produce periodograms that
are free from thruster firing signals.

\begin{figure}
\begin{center}
	\subfigure[]{
            \label{fig:1}
	    \includegraphics[width=3in]{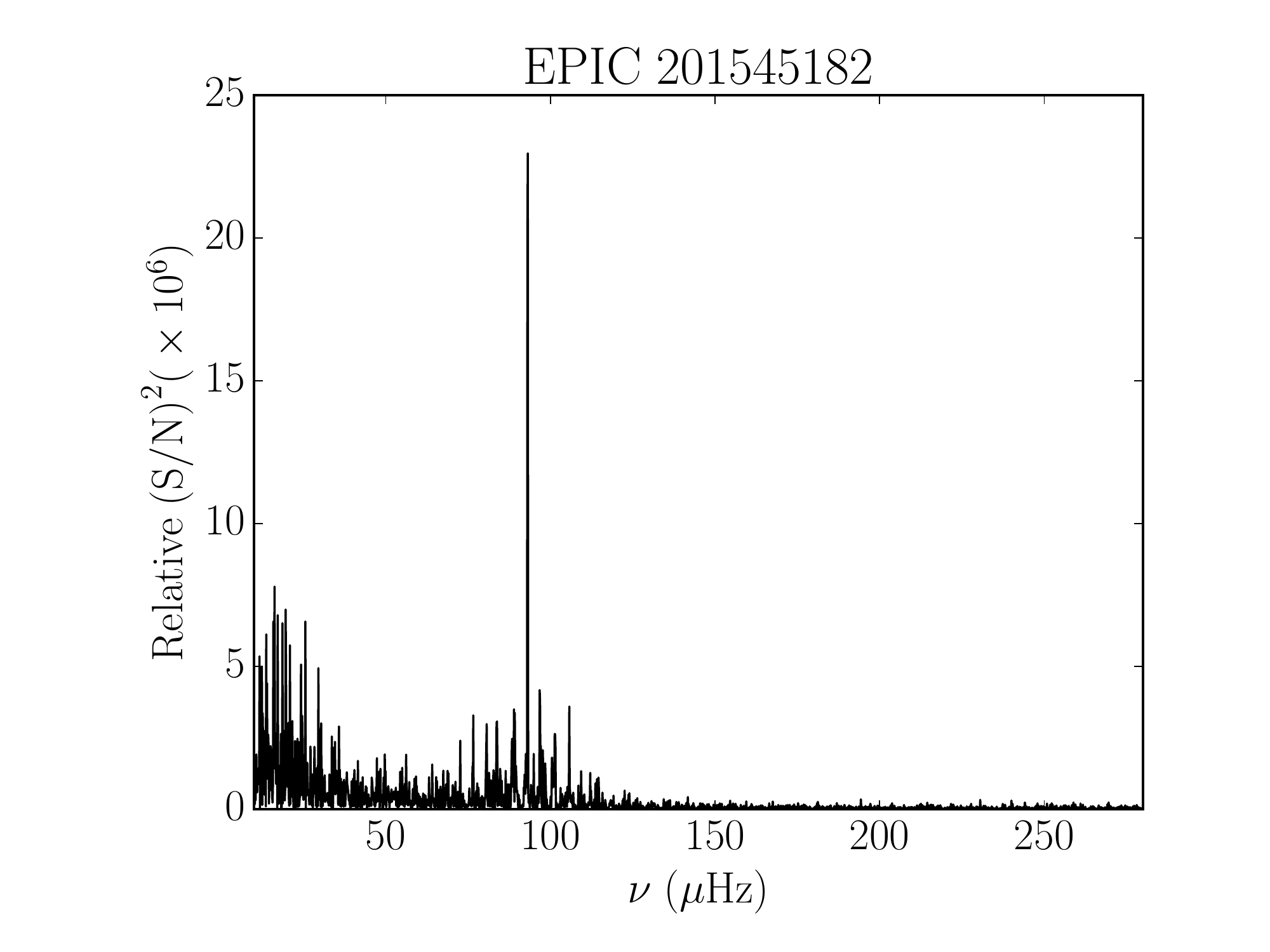}
        }
	\subfigure[]{
            \label{fig:3}
	    \includegraphics[width=3in]{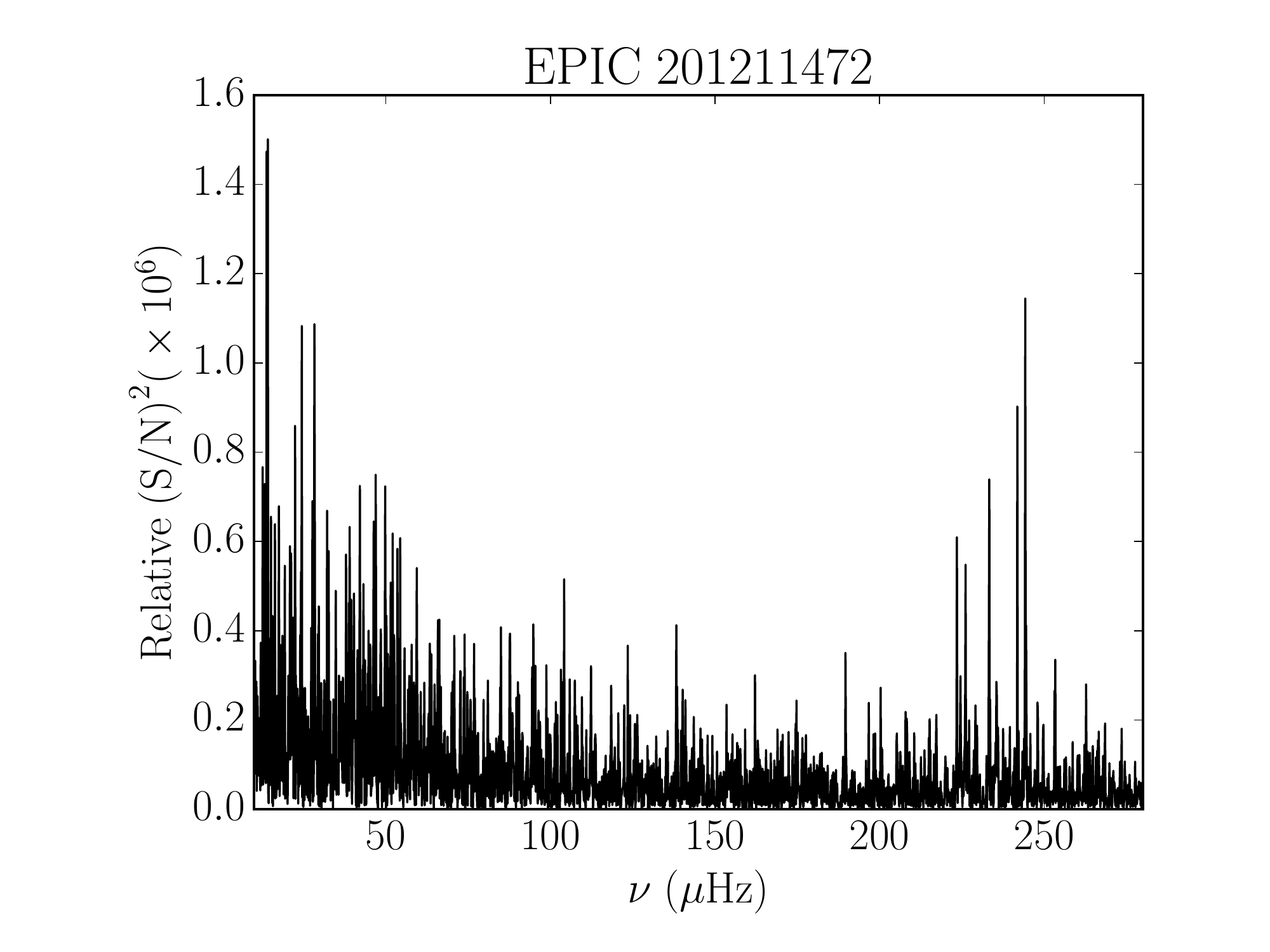}
        }
	\subfigure[]{
            \label{fig:4}
	    \includegraphics[width=3in]{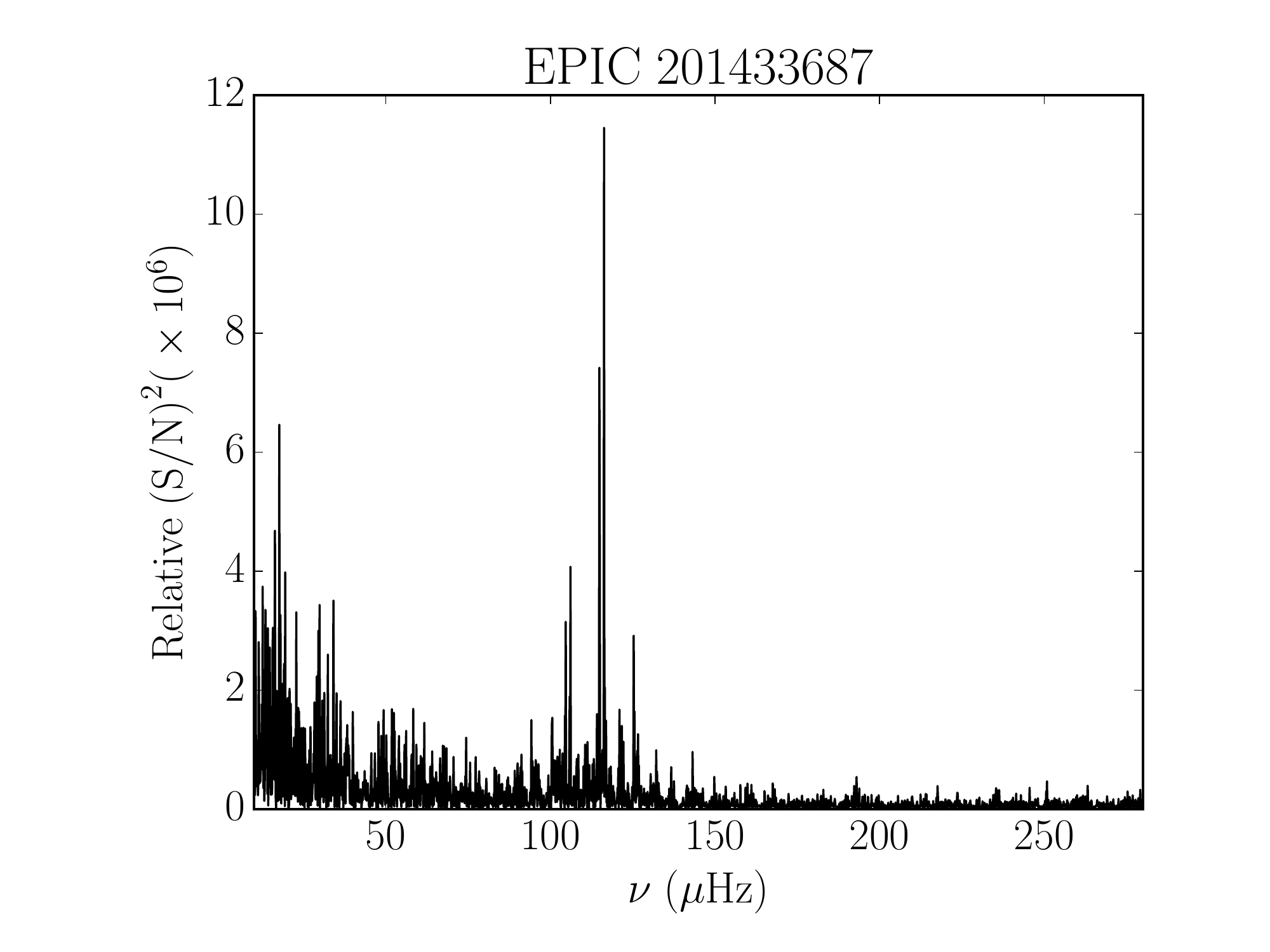}
        }
	\subfigure[]{
            \label{fig:5}
	    \includegraphics[width=3in]{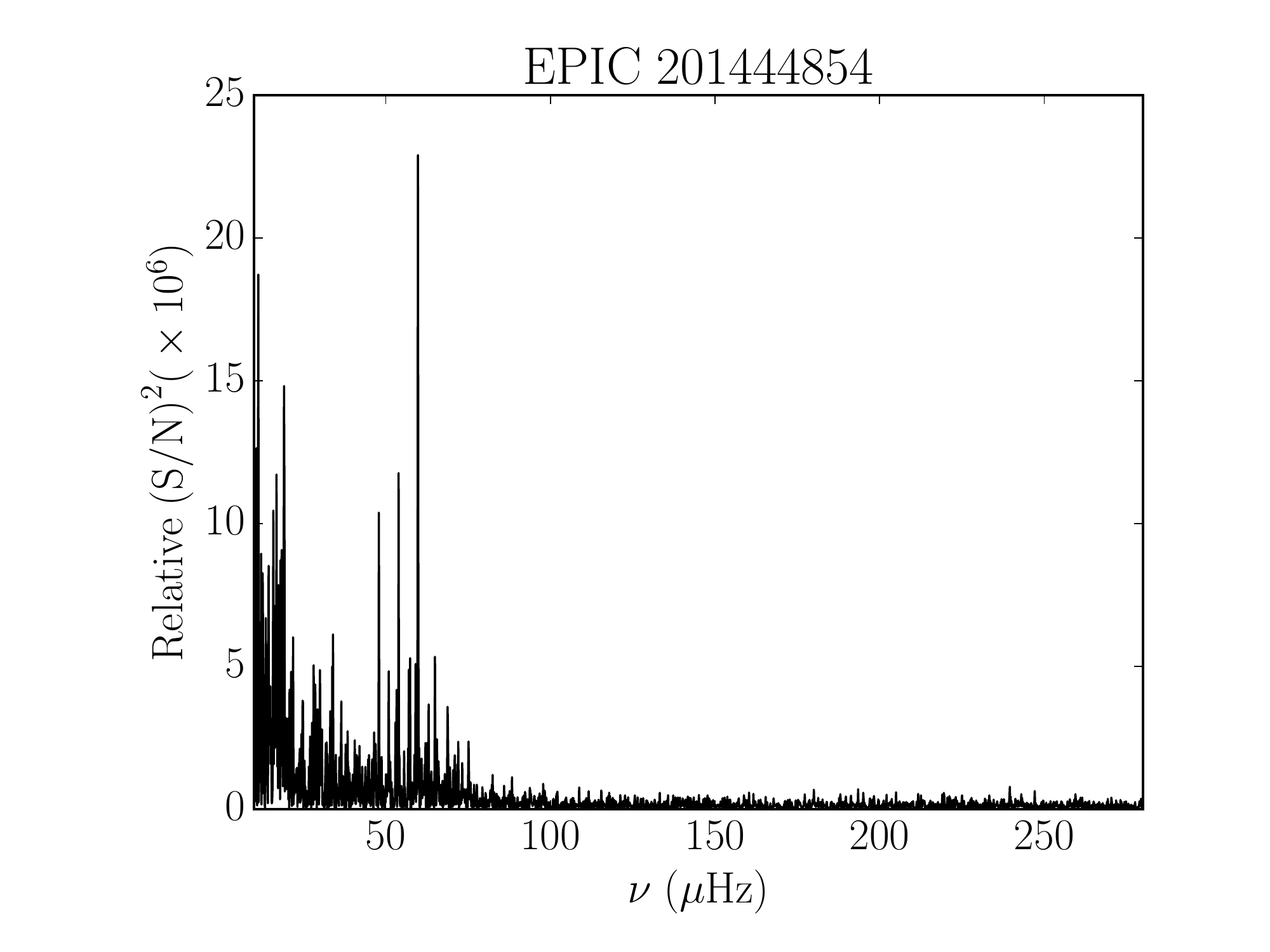}
        }
	\subfigure[]{
            \label{fig:6}
	    \includegraphics[width=3in]{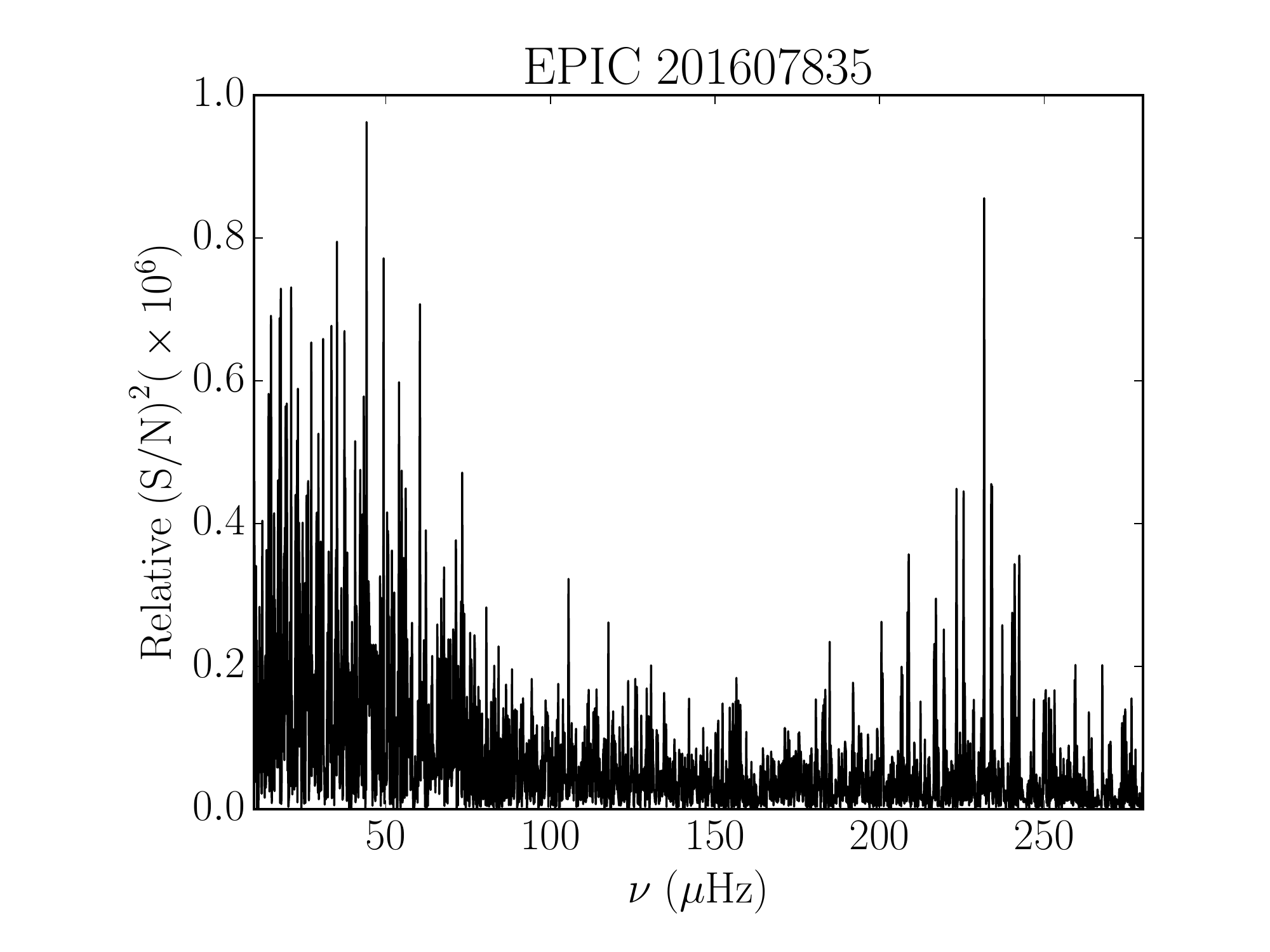}
        }
    \end{center}
    \caption{SIPs of 6 long cadence {\it K2} giants with asteroseismic
	    oscillations.
	    These were selected from the guest observing program, GO1059 and
	    identified using the method of \citet{Huber2009}.
\label{fig:astero_examples}}
\end{figure}

In order to search for high signal-to-noise asteroseismic modes in the giant
star candidates of GO1059, we searched for a power excess in the SIPs using the
method of \citet{Huber2009}: autocorrelation functions were calculated for
sections of the SIP in order to search for regions of increased
correlation and locate the frequency of maximum power.
The increased correlation arises from the even frequency spacing of acoustic
modes, and the frequency of maximum correlation at the location of the power
excess corresponds to the large frequecy separation, $\Delta\nu$.
Figures \ref{fig:1} to \ref{fig:6} show example power spectra of 6 targets for
which we detect pulsations using this method.

\begin{figure*}
\begin{center}
\includegraphics[width=6in, clip=true]{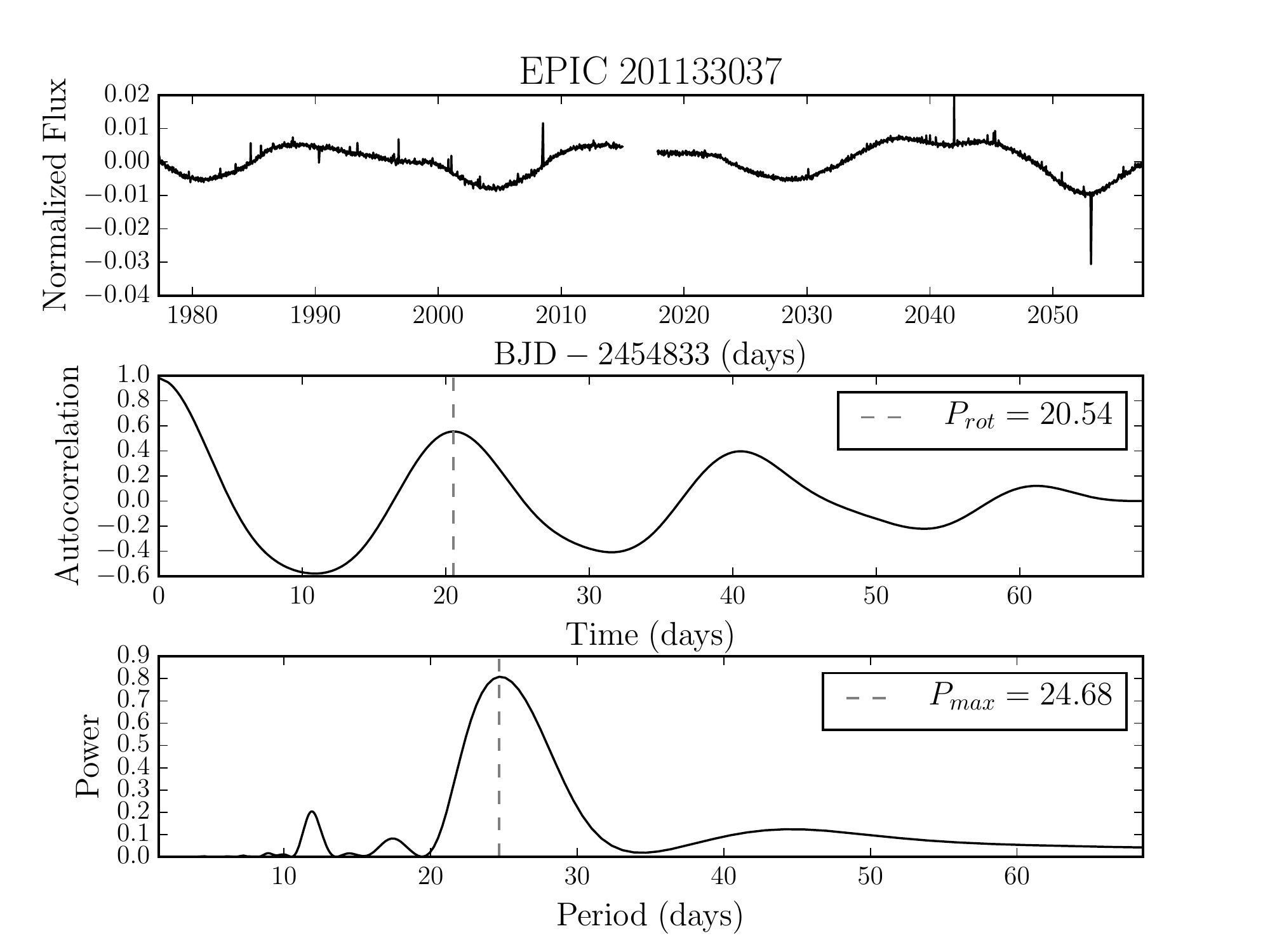}
\caption{{\it Top}: Light curve of EPIC 201133037, detrended using the method
	of VJ14.
	{\it Middle}: Autocorrelation function of the detrended light
	curve. The autocorrelation function method measures a rotation period
	of 21 days for this star.
	{\it Bottom}: The LS periodogram of the detrended light curve.
	The highest peak in the periodogram is located at 25 days.}
\label{fig:rotation_poster_child}
\end{center}
\end{figure*}

\begin{figure*}
\begin{center}
\includegraphics[width=6in, clip=true]{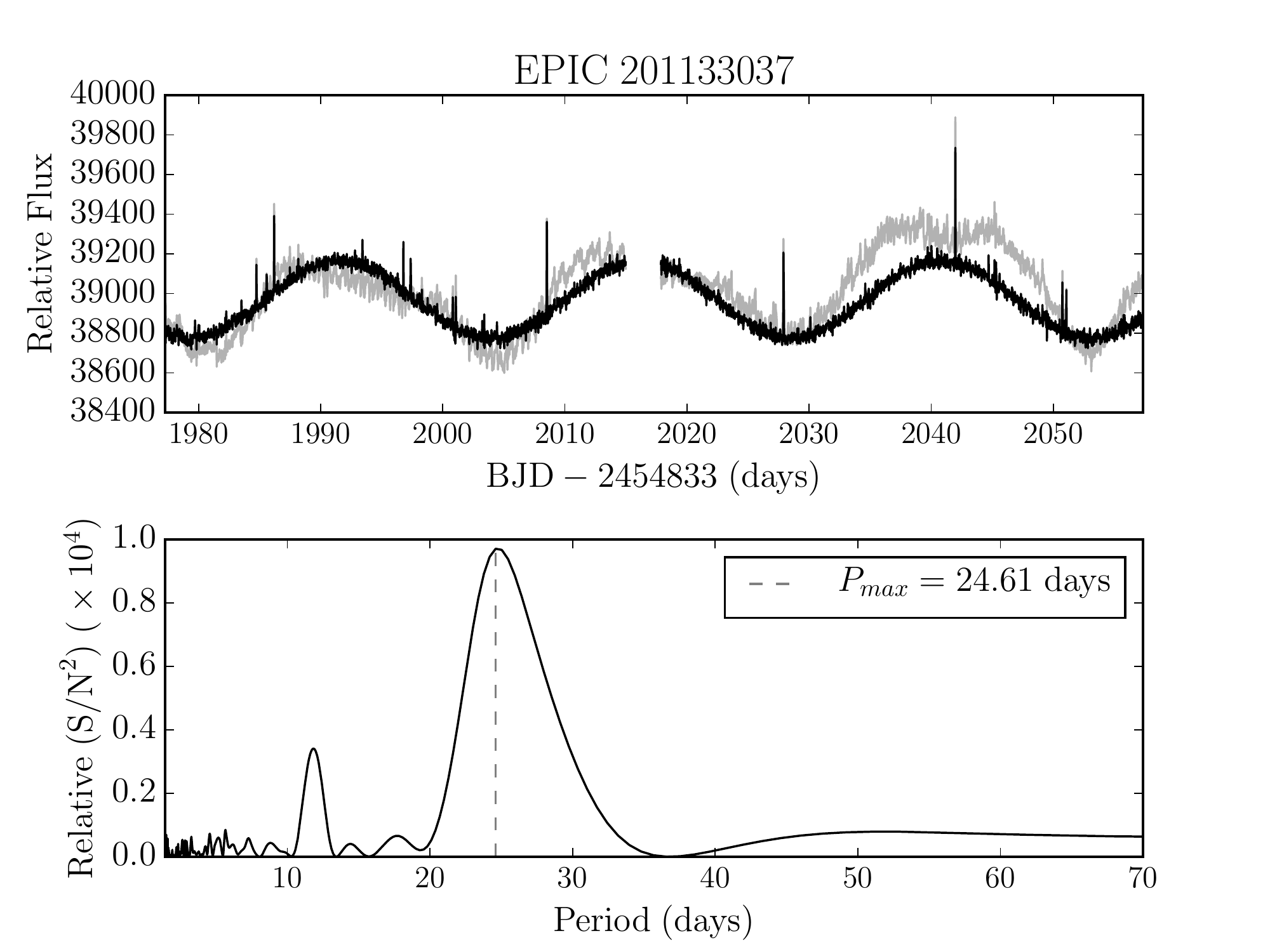}
\caption{{\it Top}: The raw light curve of EPIC \rpc\ is shown in grey and
	the conditioned light curve is shown in black. The conditioned light
	curve is produced by removing the trends that best describe the
	data, at the best fitting frequency. {\it Bottom}: An SIP of the raw
	light curve, produced by modelling the data using the top 150 ELCs
	plus a sine and cosine function at a range of frequencies.
	The highest peak in the SIP is located at 25 days.}
\label{fig:K2_rotation_poster_child}
\end{center}
\end{figure*}

The top panel of figure \ref{fig:rotation_poster_child} shows the
VJ14-detrended light curve of an active, rotating star, EPIC \rpc, with
a linear trend subtracted off.
The brightness fluctuations clearly visible in the light curve of this target
are produced by cool active regions on the stellar surface, which reduce
the stellar flux periodically.
The rotation period of this star is therefore around 20 days.
The middle and bottom panels show an ACF and LS periodogram of the
detrended light curve.
The top panel of figure \ref{fig:K2_rotation_poster_child} shows the raw light
curve of the same target in grey,  with the conditioned light curve in black.
This conditioned light curve was produced by removing the best fitting
systematic trends, described by a certain combination of the ELCs, at the best
fitting period of the sinusoid.
The bottom panel shows the SIP.

\begin{figure*}
\begin{center}
\includegraphics[width=6in, clip=true]{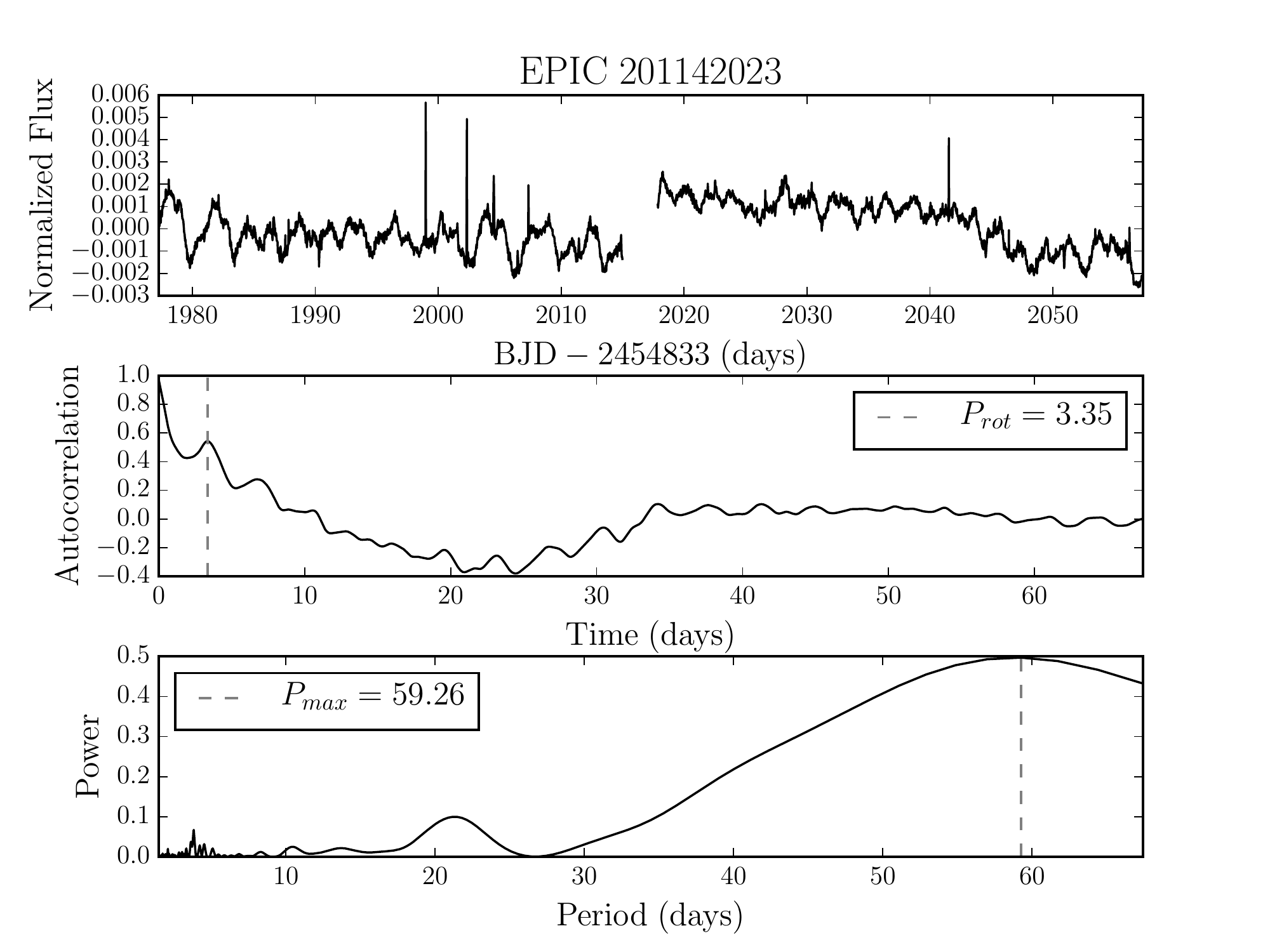}
\caption{{\it Top}: Light curve of EPIC \rpcc, detrended using the method
	of VJ14.
	{\it Middle}: Autocorrelation function of the detrended light curve.
	The autocorrelation function method measures a rotation period of 3
	days for this star.
	{\it Bottom}: The LS periodogram of the detrended light curve.
	The highest peak in the periodogram is located at 59 days and is likely
	to be a systematic trend produced by spacecraft pointing variations.}
\label{fig:rotation_poster_child2}
\end{center}
\end{figure*}

\begin{figure*}
\begin{center}
\includegraphics[width=6in, clip=true]{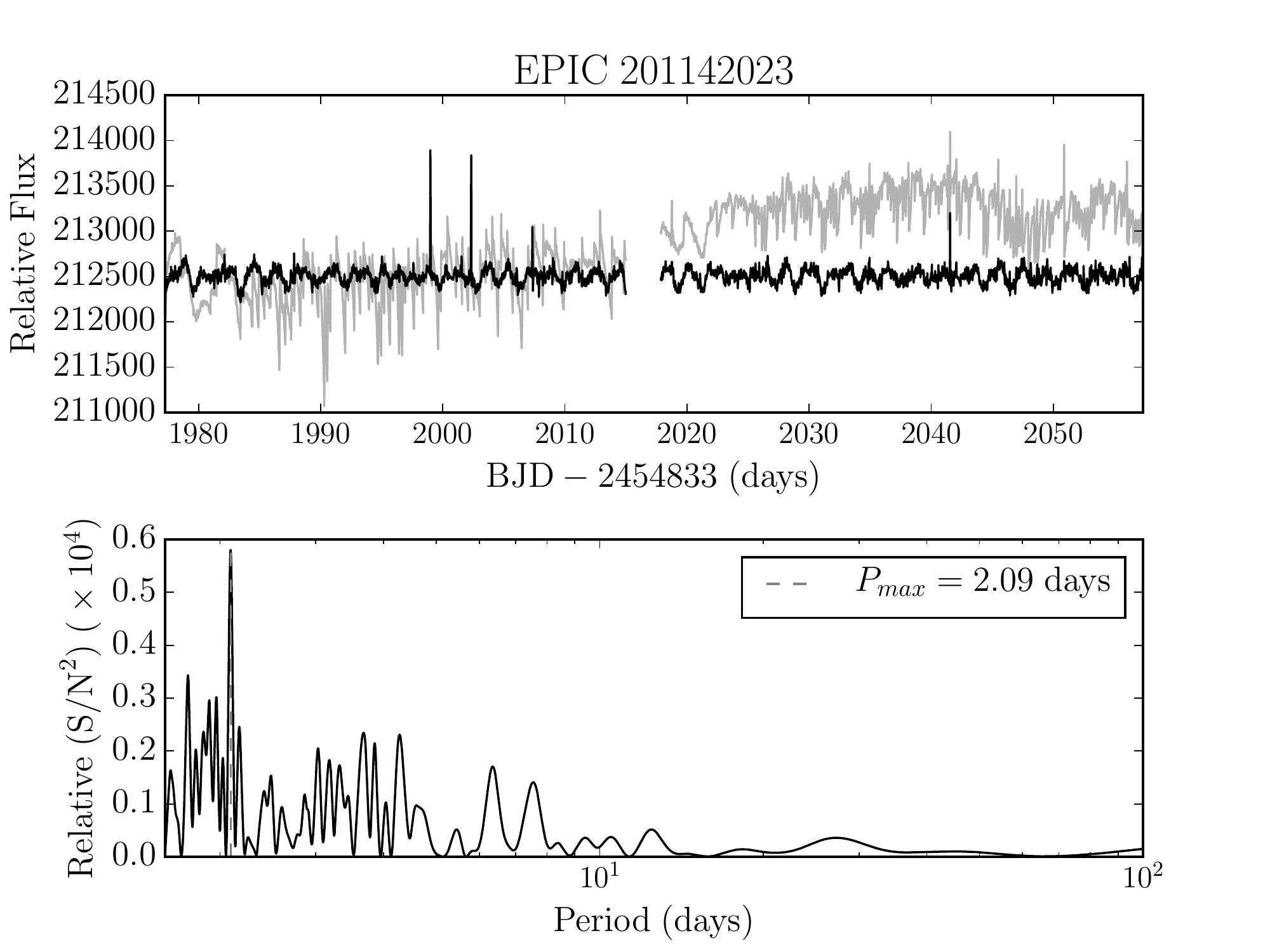}
\caption{{\it Top}: The raw light curve of EPIC \rpcc\ is shown in grey and
	the conditioned light curve is shown in black. The conditioned light
	curve is produced by removing the trends that best describe the
	data, at the best fitting frequency.
	{\it Bottom}: An SIP, produced by modelling the data as a linear
	combination of the top 150 ELCs plus a sine and cosine function at a
	range of frequencies, measuring a rotation period of 2 days.
	The SIP is, by definition, insensitive to the long-timescale
	systematics that dominate the LS periodogram of the detrended data,
	shown in figure \ref{fig:rotation_poster_child2}.}
\label{fig:K2_rotation_poster_child2}
\end{center}
\end{figure*}

Each of these three methods measures a rotation period of around 20 days for
this target.
This example demonstates the ability of the SIP to recover rotation periods
that agree with those measured from detrended light curves by autocorrelation.
We also include an example that demonstrates the ability of the SIP to
outperform a periodogram of detrended data.
Figure \ref{fig:rotation_poster_child2} shows the light curve, ACF and LS
periodogram of another rotating star, EPIC \rpcc\ and figure
\ref{fig:K2_rotation_poster_child2} shows its SIP.
This star shows lower amplitude variability than the previous example and the
careful treatment of systematics is much more important.
Whereas the ACF method is able to measure a rotation of $\sim$ 3 days for
this star, the LS periodogram of the detrended light curve incorrectly
measures a period of 59 days.
Although there is a small peak at the rotation period of the star, it is not
the dominant periodic signal.
The SIP method is, by definition, insensitive to these long-term systematics
and is able to measure a period of $\sim$ 2 days.
This example further demonstrates the fact that long-term systematic trends
caused by slow pointing variations are often not removed by conventional
detrending methods.
The 59 day signal is almost certainly a systematic trend and not an
astrophysical signal because it does not appear in the SIP.
It is well described by the ELCs and must therefore be common to many stars.

We have shown that the SIP method is able to measure stellar rotation periods
and does better than producing periodograms from detrended data.
However, it has been shown that the ACF method often performs better than
periodogram methods in general for measuring stellar rotation periods
\citep[][]{McQuillan2013, McQuillan2013b, Mazeh2015}.
For stars with relatively high-amplitude variability, for which perfect
removal of systematic trends is less important, performing the ACF method
on detrended data is likely to produce similar results to the SIP method.
The SIP method is ideally suited to low-amplitude cases, where systematic
trends could drastically influence rotation period measurements.
Whilst the SIP method may outperform ACF in the low-amplitude cases,
any `marginal' rotation period measurements calculated using either method
should be treated with caution unless a representative uncertainty is provided.
In general neither ACF nor periodogram methods are equipped to provide
such uncertainties.
In practise, we recommend using both the SIP and ACF methods, in combination
with a by-eye check, to measure rotation periods for {\it K2} stars.

\subsection{Tests and discussion}

\begin{figure*}[h]
\begin{center}
\includegraphics[width=6in, clip=true]{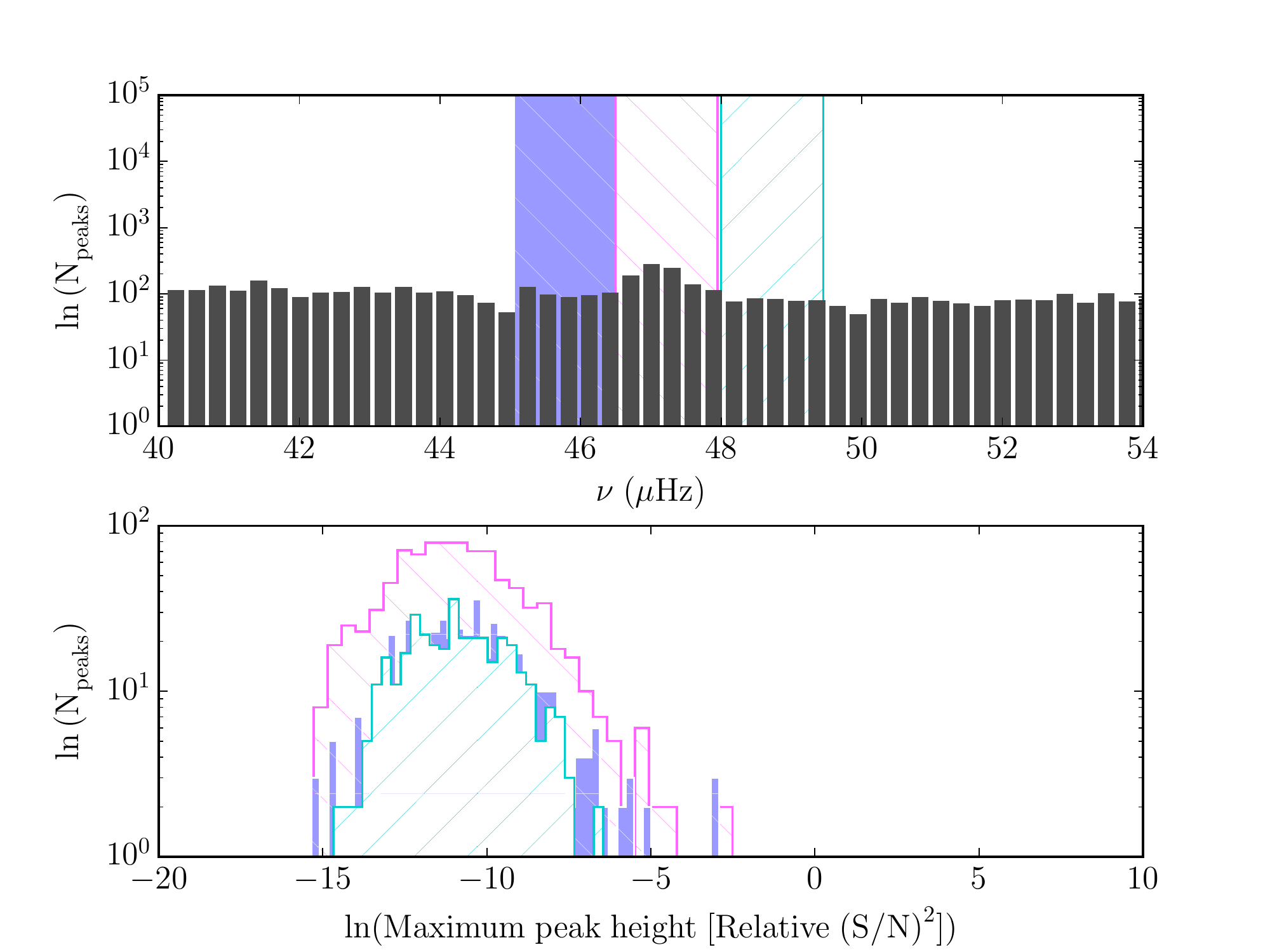}
\caption{{\it Top:} Histogram of the frequencies of the highest peaks in the
	SIPs of \nGO\ {\it K2} targets within the range 40--54 \uHz.
	{\it Bottom:} Histograms of peak heights within the correspondingly
	colored ranges indicated in the top panel.
	Whilst there is a larger number maximum peaks around 47 \uHz\ (the
	frequency corresponding to the 6 hour thruster fire) the amplitudes of
	these maximum peaks are comparable to the maximum peak heights just
	above and just below this frequency.}
\label{fig:sip_hist}
\end{center}
\end{figure*}

\begin{figure*}[h]
\begin{center}
\includegraphics[width=6in, clip=true]{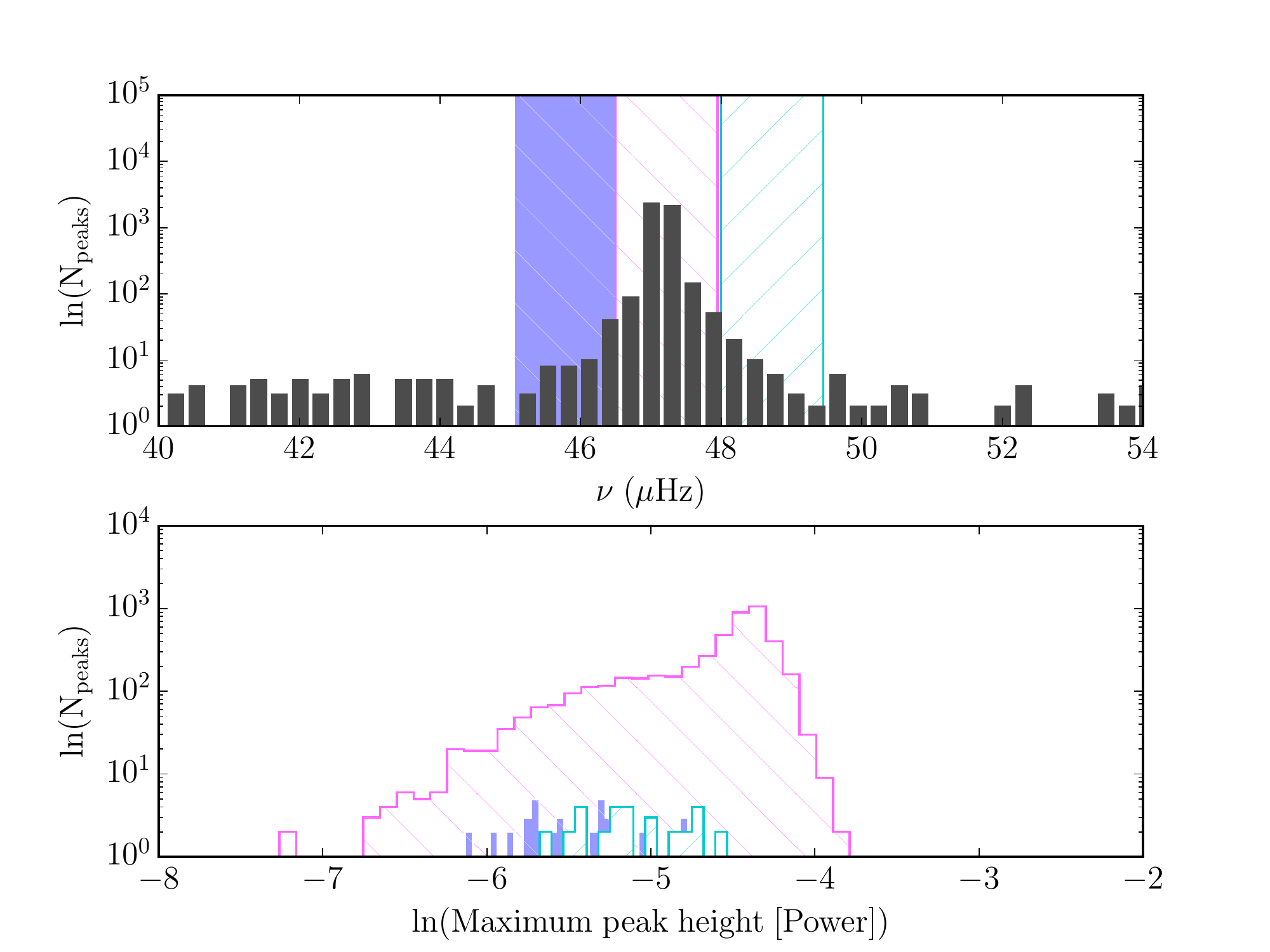}
\caption{{\it Top:} Histogram of the frequencies of the highest peaks in the
	LS periodograms of the \citet{Vanderburg2014} light curves of \nGO\
	{\it K2} targets within the range 40--54 \uHz.
	{\it Bottom:} Histograms of peak heights within the correspondingly
	colored ranges indicated in the top panel.
	The frequency of maximum peak height was $\sim$ 47 \uHz\ in almost
	every periodogram.
	Furthermore, the distribution of maximum peak height within the range
	46.5-48 \uHz\ is skewed towards higher powers, i.e. a large fraction of
	the peaks at $\sim$ 47 \uHz\ have a large power.
}
\label{fig:vbg_hist}
\end{center}
\end{figure*}

In order to demonstrate the consistent ability of the SIP method
to remove the signal at 47\uHz, corresponding to the periodic $\sim$6 hour
thruster firings, we computed SIPs for \nGO\ targets from the GO1049
proposal: ``Galactic Archaeology on a grand scale" (PI: Stello, D.).
For each target, an SIP of its raw photometry and a LS periodogram of its
VJ14 light curve was calculated for frequencies between
40 and 54 \uHz.
Both the height and frequency of the highest peak in the SIP and the highest
peak in the LS periodogram were recorded.
A histogram of the frequencies of the highest peaks in the SIPs of all \nGO\
targets is shown in the top panel of figure \ref{fig:sip_hist}.
The bottom panel shows the histograms of peak heights within the
correspondingly colored ranges indicated in the top panel.
This figure shows that while there are a greater number of maximum peaks
around 47 \uHz, the S/Ns of these peaks are comparable to those found just
above and just below this frequency.
Figure \ref{fig:vbg_hist} shows the equivalent results for the
VJ14 light curves.
There is a significant number of large peaks at $\sim$47 \uHz\ in the LS
periodograms of the detrended light curves; the highest peak in the LS
periodograms was almost always located at $\sim$ 47 \uHz.
Furthermore, the distribution of peak power within the range 46.5-48 \uHz\ is
skewed towards higher powers, i.e. a substantial fraction of the peaks at
$\sim$ 47 \uHz\ have a large power.
The SIP method is able to consistently remove the 47 \uHz\ signal which is
present in almost every VJ14 light curve.

\begin{figure}
\begin{center}
\includegraphics[width=6in, clip=true]{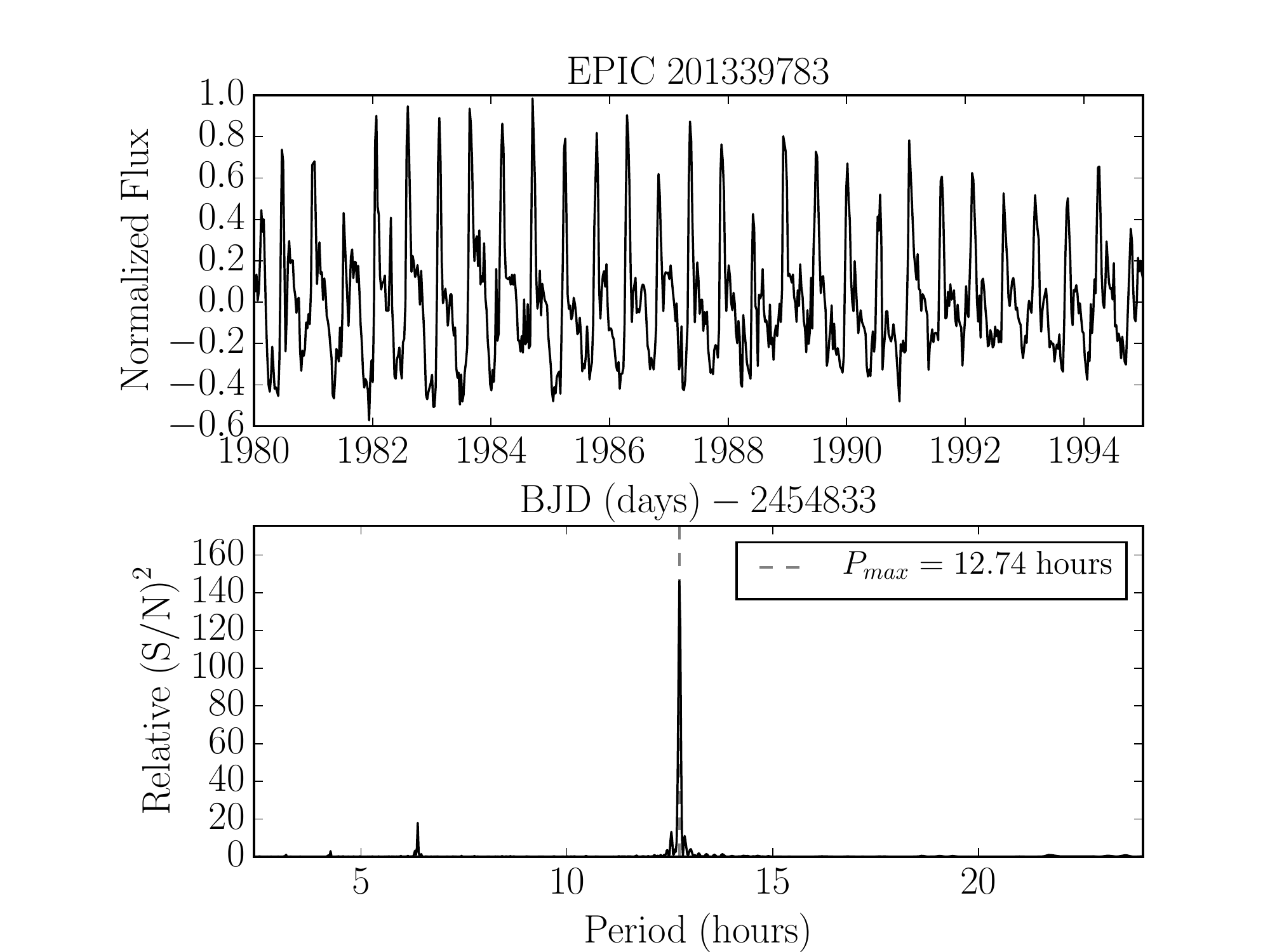}
\caption{{\it Top:} The light curve of RR Lyrae star, EPIC 201339783,
	conditioned on the highest amplitude sinusoidal signal found in the
	SIP. {\it Bottom:} The systematic-insensitive periodogram of
	this light curve. This target was selected from GO1018
	(PI: Plachy, E.).}
\label{fig:RRLyrae}
\end{center}
\end{figure}

\begin{figure*}
\begin{center}
\includegraphics[width=6in, clip=true]{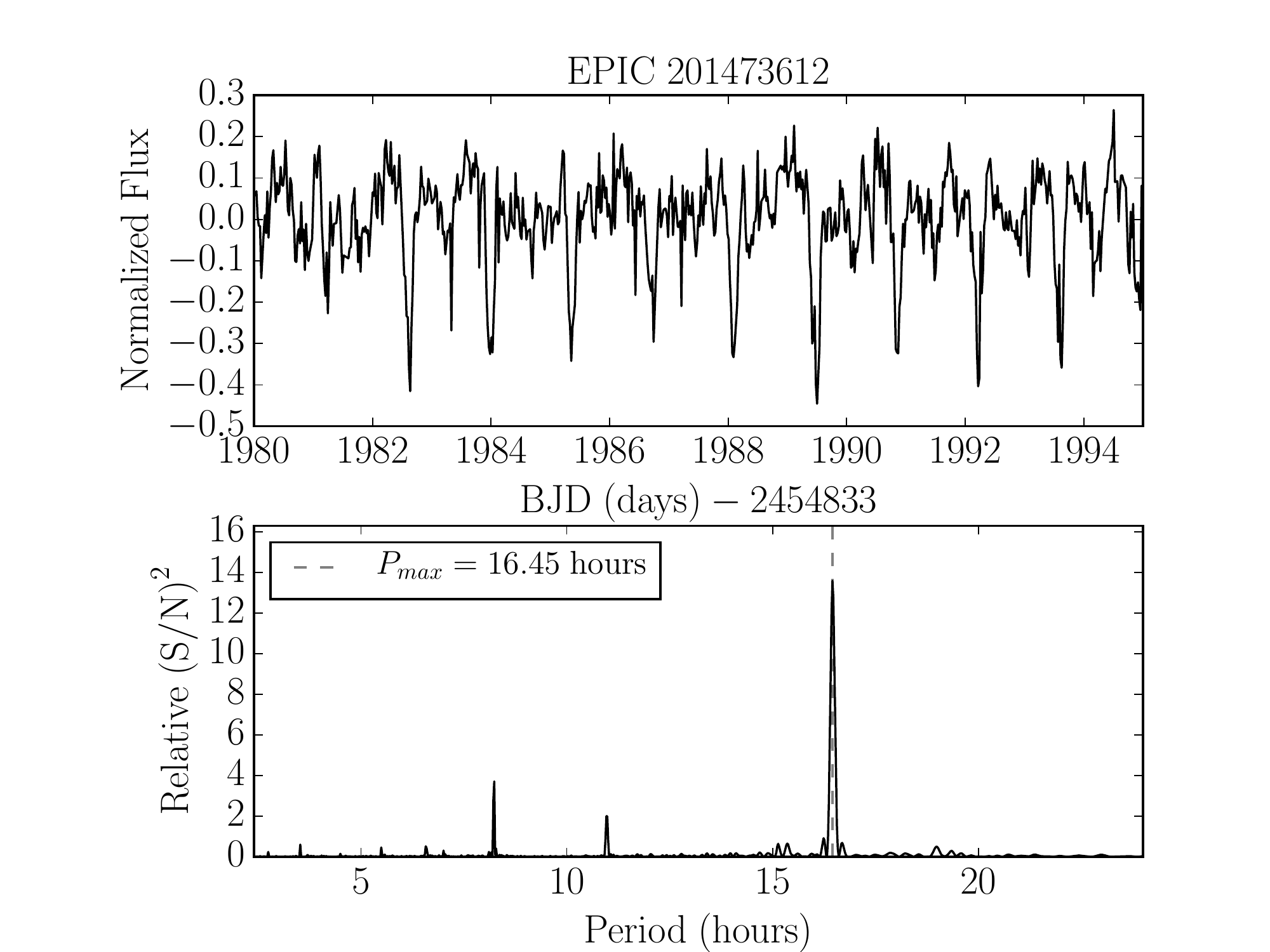}
\caption{{\it Top:} The light curve of eclipsing binary, EPIC 201473612,
	conditioned on the highest amplitude sinusoidal signal found in the
	SIP. {\it Bottom:} The systematic-insensitive periodogram of
	this light curve. This target was selected from the catalogue of EBs
	and variable stars of \citet{Armstrong2015}.}
\label{fig:EB}
\end{center}
\end{figure*}

\begin{figure*}
\begin{center}
\includegraphics[width=6in, clip=true]{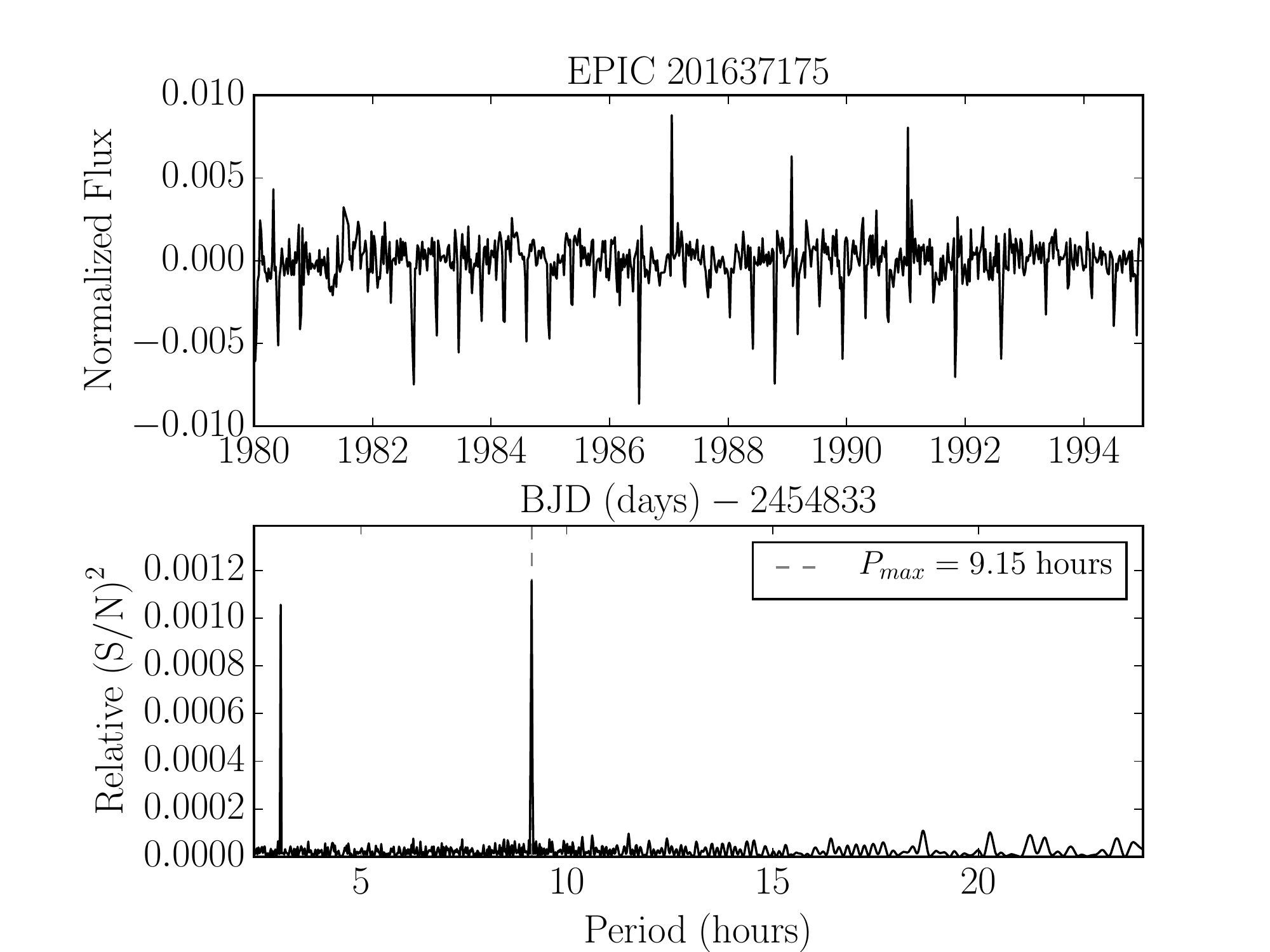}
\caption{{\it Top:} The light curve of exoplanet candidate, EPIC 201637175,
	conditioned on the highest amplitude sinusoidal signal found in the
	periodogram. {\it Bottom:} The systematic-insensitive periodogram of
	this target.}
\label{fig:planet}
\end{center}
\end{figure*}

Figures \ref{fig:RRLyrae} and \ref{fig:EB} show the conditioned light curves
and SIPs of an RR Lyrae star, selected from Guest observer program GO1018;
``Long Cadence RR Lyrae targets" (PI: Plachy, E.) and
an Eclipsing Binary (EB), selected from the \citet{Armstrong2015} {\it K2} EB
and variable star catalogue respectively.
Figure \ref{fig:planet} shows the conditioned light curve and SIP of the
short-period, disintegrating planet candidate discovered by
\citet{Sanchis-Ojeda2015}.
This planet has a period of around 9 hours, short enough to be detectable
with a periodogram, as was demonstrated for a number of ultra-short
period {\it Kepler} exoplanets by \citet{Sanchis-Ojeda2014}.
The top panels of these three figures show the {\it K2} light curves of these
objects, conditioned on the highest S/N sinusoidal signal in the
periodograms.
These light curves were produced by subtracting the trends that best describe
the data at the highest S/N period found in the SIP.

Photometric variability in dwarf stars on timescales less than 8 hours, often
known as flicker, has been linked to surface gravity
\citep[][]{Bastien2013, Kipping2014}.
The metrics used to quantify photometric variability include finding the range
in intensity, counting the number of zero crossings and calculating the
root-mean-square (RMS) of the light curve.
Although these features are related to signal processing, they are operations
performed on detrended light curves, not inferred from periodograms.
However, it may be possible to derive a property of the periodogram that scales
with the density or surface gravity of a star, for example, the mean excess
power at frequencies near those relevent to granulation timescales.
The SIP method presented here would be useful for such a technique.

\section{Conclusions}
\label{sec:conclusions}
We demonstrate that modelling campaign 1 {\it K2} photometry as a linear
combination of 150 PCA components plus a sinusoid can produce periodograms
that are almost completely free from instrumental systematic signals, without
the need for detrending.
We find that the 47 $\mu$Hz signal, generated by the spacecraft thruster
fires is not present in the vast majority of Systematics-Insensitive
Periodograms (SIPs) for more than 4000 targets
selected from the {\it K2} guest observer program, GO1059, ``Galactic
Archaeology on a grand scale" (PI: Stello, D.).
The SIP method is highly successful for campaign 1 targets where the large
number of stars, observed for a baseline of 80 days ensures that most of the
systematics are captured in the ELCs and we anticipate that it will
be equally effective for the up-and-coming campaigns.

The SIP method is capable of detecting periodicities in {\it K2} data in the
region of frequency space relevent to the study of asteroseismic oscillations
in giant stars and for any signals with a timescale close to 6 hours.
It is also effective at measuring stellar rotation periods and is an
improvement upon a simple LS periodogram of detrended data.
In practise, the best approach for measuring rotation periods in {\it K2} data
is likely to be a combination of the SIP method and the ACF method, where
autocorrelation is performed on detrended light curves.
The SIP code is available for public use and can be found at
\url{https://github.com/RuthAngus/SIPK2}.

It is a pleasure to thank Dan Huber (Sydney) who provided many excellent
comments for this paper and useful asteroseismology tips.
We would also like to thank David Hogg (NYU) and Suzanne Aigrain (Oxford)
for their comments, plus Andrew Vanderburg (Harvard), Ben Montet (Caltech)
and Stephanie Douglas (Columbia) for their extremely helpful suggestions and recommendations regarding this project.
J.A.J is supported by generous grants from the David and Lucile Packard and
Alfred P. Sloan Foundations.
The data presented in this paper were obtained from the Mikulski Archive for
Space Telescopes (MAST).
STScI is operated by the Association of Universities for Research in Astronomy,
Inc., under NASA contract NAS5-26555.
Support for MAST for non-HST data is provided by the NASA Office of Space
Science via grant NNX09AF08G and by other grants and contracts.
This paper includes data collected by the Kepler mission.
Funding for the Kepler mission is provided by the NASA Science Mission
directorate.

\bibliographystyle{plainnat}
\bibliography{k2rotation}
\end{document}